# The X-ray Spectrum Of The Black Hole Candidate Swift J1753.5-0127

Reham Mostafa

Fayoum University, Faculty of Science, Physics Department, PO Box 63514, Fayoum, Egypt

reham.m.ayoub@gmail.com

Mariano Mendez, Beike Hiemstra, Paolo Soleri

Kapteyn Astronomical Institute, University of Groningen, PO Box 800, 9700 AV Groningen, The Netherlands

mariano@astro.rug.nl, hiemstra@astro.rug.nl, and soleri@astro.rug.nl

Tomaso Belloni,

INAF-Osservatorio Astronomico di Brera, Via E. Bianchi 46, I-23807, Merate (LC), Italy

tomaso.belloni@brera.inaf.it

Alaa I. Ibrahim

Massachusetts Institute of Technology, Kalvi Institute for Astrophysics and Space Research, Cambridge, MA 02139, U.S.A.

ai@space.mit.edu

and

Mohammed N. Yasein

Fayoum University, Faculty of Science, Physics Department, PO Box 63514, Fayoum, Egypt

mkb00@fayoum.edu.eg

**ABSTRACT**

We present a spectral analysis of the black hole candidate and X-ray transient source Swift J1753.5 0127 making use of simultaneous observations of *XMM-Newton* and *Rossi X-ray Timing Explorer* (*RXTE*) in 2006, when the source was in outburst. The aim of this paper is to test whether a thermal component due to the accretion disc is present in the X-ray spectrum. We fit the data with a range of spectral models, and we find that for all of these models the fits to the X-ray energy spectra significantly require the addition of the disc black-body component. We also find a broad iron emission line at around 6.5 keV, most likely due to iron in the accretion disc. Our results confirm the existence of a cool inner disc extending near or close to the innermost circular orbit (ISCO).We further discovered broad emission lines of NVII and OVIII at ~ 0.52 keV and 0.65 keV, respectively in the RGS spectrum of Swift J1753.5-0127.

## 1. INTRODUCTION

Swift J1753.5-0127 is a stellar-mass black hole candidate which was discovered with *Swift*/BAT on 2005 May 30 (Palmer et al. 2005). The source was observed in the low/hard state (LHS) and has remained in such a state during the whole outburst (Cadolle Bel et al. 2007, Zhang et al. 2007). Spectral and timing analysis made by *Swift*/XRT and *RXTE*/PCA indicate a hard power-law spectrum and a 0.6 Hz quasi periodic oscillation (Morris et al. 2005, Morgan et al. 2005). A double-peaked Hα emission line was found by optical spectroscopy of Swift J1753.5-0127 (Torres et al. 2005). Recently, Hiemstra et al. (2009) discovered a broad iron emission line between 6.0 and 7.0 keV; this line was later on independently confirmed by Reis et al. (2009).

Black hole binaries (BHBs) exhibit different X-ray states, denoted as low/hard(or hard), high/soft (or thermal), and very hard (or steep power law) states. In the low/hard state, the X-ray spectrum of a black hole is dominated by a hard power-law component (> 80% of the flux in the 2.0 – 20.0 keV band) with photon index in the range $1.5 < \Gamma < 2.1$ and this component is explained to be due to the Comptonization of soft photons by thermal plasma of electrons, whereas the high/soft state is characterized by a thermal component that contributes > 75% of the total 2.0 – 20.0 KeV unabsorbed flux. On the other hand, in the steep power law state the X-ray spectrum is much steeper, with $\Gamma > 2.4$ and also show a thermal component (McClintock & Remillard 2006). It is believed that BHB usually enters the low/hard state in the early and late phases of an outburst. The disc in the low/hard state may be truncated at some large radius, whereas in the high/soft state it extends almost in to the last stable orbit. The observed hard X-rays are formed by inverse Compton scattering of soft photons from the disc. The question here is the structure of the emitting region. There are two classes of model which explain the geometry of the emitting region in the low/hard state: 'truncated disc' model and 'corona' model. The truncated disc model was early proposed by Shapiro et al. (1976). It assumes a cool, optically thick, geometrically thin disc that truncated at large radii and hot, optically thin, geometrically thick flows at small radii which produces the hard X-rays. The reflection spectrum and Fe fluorescence result from the interaction of the hard X-ray by the inner part of the truncated disc. When cooling of the accretion flow is not efficient, then most of the viscously dissipated energy is advected radially with the flow rather than being radiated (Narayan & Yi 1994) and such flow called advection-dominated accretion flow (ADAF). The corona model involved a corona of hot, optically thin plasma covers a cold, optically thick accretion disc which provides the input of soft photons for comptonization (Bisnovatyi-Kogan & Blinnikov 1977; Liang & Price 1977). It was suggested that hot corona is formed as a result of magnetorotational instabilities in the disc and the buoyancy of the generated magnetic field (Tout & Pringle 1992; Beloborodov 2001). The geometry of such hot corona could be either sandwich geometry (slab-corona model) in which it covers most of the cold disc (Haardt & Maraschi 1991, 1993), or patchy corona (Galeev et al. 1979, Haardt et al. 1994) in which a number of small coronal regions sit above the cold accretion disc. In both geometries the disc extends down close to the ISCO. However, such model could not explain the observations of some black hole sources in the hard state (e.g., Gierlinski et al. 1997). Beloborodov (1999) pointed out that due to the anisotropy of the energy dissipation process and due to radiation pressure from the disc the flaring plasma is likely to be outflow away from the disc with mildly relativistic bulk velocities.

In this work we reanalyze the same data of the BHB Swift J1753.5-0127 used by Miller, Homan & Miniutti (2006a) and Hiemstra et al. (2009). We applied different models with and without the disc blackbody component to test the presence of the thermal component in the LHS.

## 2. OBSERVATIONS AND DATA REDUCTION

In this paper we analyzed the observations of Swift J1753.5-0127 which was observed simultaneously with *XMM-Newton* and *RXTE* on 2006 March 24, about 270 days after the start of the outburst, at 16:00:31 UT and 17:31:11 UT, respectively. Let us detail how we reduced the data from both *RXTE* and *XMM-Newton* telescopes.

### 2.1 RXTE Data Reduction

We obtained the *RXTE* data from both PCA and HEXTE detectors for the observation ID 92082-02-01-00. For both detectors we reduced our data using the HEASOFT v 6.3 software package. For PCA the exposure time was 2.3 ksec. Here we used 'Standard 2' mode data from all Xenon layers of only PCU-2. We created the source spectrum. Thereafter, we used the FTOOL 'pcabackest' to generate the background file. In our work, we used the 'brightsource' background model (pca-bkgd-cmbright-e5cv2005) to extract the background spectrum. We made the response matrix using the task 'pcarsp'. Also, we added 0.6% systematic errors to the PCA spectrum (Miller, Homan & Miniutti 2006a) using the FTOOL 'grppha'. For HEXTE, we reduced 'Archive mode' data from only HEXTE cluster-B. The exposure time was 0.78 ksec. We extracted the background and source spectra. In addition, we produced the response matrix using the FTOOL 'hxtrsp'.

*2.2 XMM-Newton Data Reduction*

We reduced the *XMM-Newton* Observational Data File (ODF) using the *XMM-Newton* Science Analysis System (SAS) version 9.0.0. To analyse our data we used the latest current calibration files (CCF). After preparing our data, we reprocessed the observation data files to get the calibrated photon event files for the RGS and EPIC-pn instruments, covering an energy range from 0.15 KeV to 15 KeV. This is done through running the pipeline processing meta-tasks *epproc* and *rgsproc*, respectively.

The SAS task *rgsproc* produced the spectra and response matrices of the first and second order of the two RGS. The observation time was 39.7 ksec for RGS1 and 39.5 ksec for RGS2. To check if the RGS data were affected by flares we created the light curve for the background. There appeared flares and so we filtered the data by creating Good Time Interval table (GTIs). Then we reprocessed the data with *rgsproc*. From the event files, we found that the EPIC-pn camera was operated in 'timing' mode. The net exposure of the EPIC cameras was 40.11 ksec. To correct for a Charge Transfer Inefficiency (CTI) effect which has been seen in EPIC-pn fast mode (timing and burst mode) data we used the SAS task *epfast* on the PN event files. By excluding bad pixels from EPIC data and then plotting the light curve, this revealed that there were no background flares, so we took the whole duration to extract our spectral data. Using the SAS task *xmmselect*, the source and background spectra were extracted by selecting the region in RAWX between 30–45 and 10–25, respectively and grouping PI channels 0–20479 by a factor of 5. Next, we created ancillary response files (arfs) and redistribution matrix file (rmfs) using the tools *arfgen* and *rmfgen*, respectively. We used *Pharbn*[1] to group the PN data to oversample the PN resolution by a factor of 3, and ensuring that each bin contains a minimum of 20 counts.

### 3. SPECTRAL ANAALYSIS AND RESULTS

This paper describes a similar analysis as the one presented in Hiemstra et al. (2009) but we do not use MOS2, since it caused some problems and gave a bad fit, and we use RGS data. We combined the spectra obtained from the four instruments PCA, HEXTE, PN, and RGS and performed spectral fits to the joint spectrum using XSPEC version 12.5.1 (Arnaud 1996). For these fits, we used PCA, HEXTE, PN, and RGS1,2 data in the 3.0–20.0 keV, 20.0–100.0 keV, 0.6–10.0 keV, and 0.4-2.1 keV energy bands, respectively. Due to the differences in the overall normalizations between instruments, we included a multiplicative constant in the model. When fitting the models, we tied all parameters except the multiplicative constants involving different instruments. Error measurements correspond to 1σ confidence intervals.

To account for the interstellar absorption along the line of sight to Swift J1753.5-0127, we included the model component PHABS in XSPEC, with cross sections from Verner et al. (1996) and abundances from Wilms, Allen & Mccray (2000). We first combined the spectra obtained from PCA, HEXTE, and PN and fitted with the simple power-law model (POWERLAW in XSPEC) to account for the hard component. We found that there are negative residuals in the fits of the PN data near 0.9 keV and 1.8 keV and we suspect that they are most likely due to systematic uncertainties in the calibration. Thus, in all fits we ignored the range 0.85–0.95 KeV and 1.7–1.9 KeV.

---

[1] http://virgo.bitp.kiev.ua/docs/xmm_sas/Pawel/reduction/pharbn

Moreover, positive residuals could be seen close to the 6 ~ 7 KeV range in the X-ray continuum, as illustrated in Fig. 1, where an emission line due to iron might be expected to be present, which was first reported by Hiemstra et al. (2009). To accommodate the emission feature, we initially modeled it with a Gaussian emission line (GAUSS in XSPEC) and then with a LAOR line profile (Laor 1991) (LAOR in XSPEC), which is thought to be produced by fluorescence when hard X-rays reflect off optically thick disc material. We found that the LAOR line was more successful at describing the broad emission-line feature. Therefore, for all fits below, the iron emission line is modeled by a LAOR line with the line energy restricted to range between 6.4 and 7 keV, the outer radius fixed at 400 Rg and the rest of the parameters free to vary.

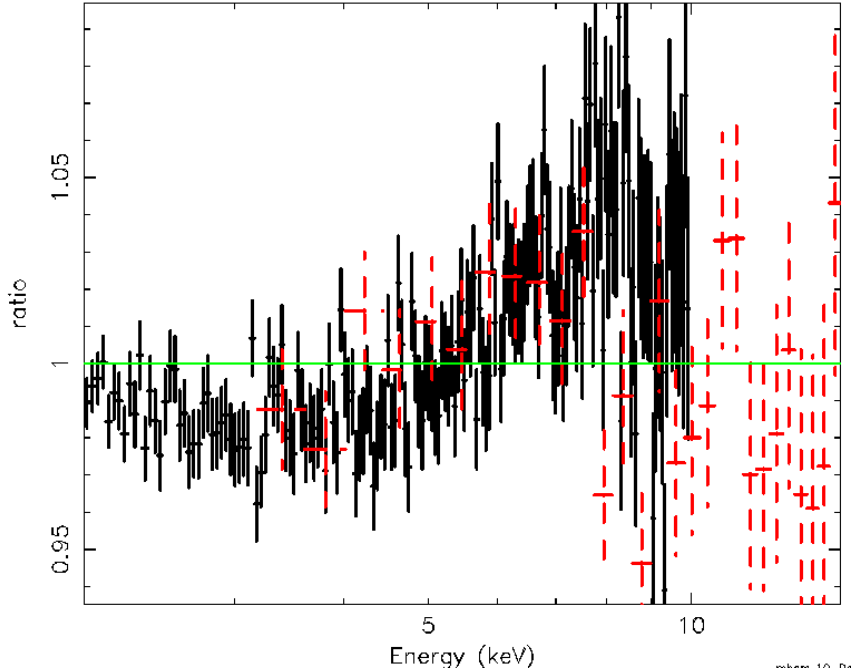

Fig.1 Broad iron line detected in the X-ray continuum of the combined spectra from PN, PCA, and HEXTE modeled by the simple power-law model.

We then combined the spectra from RGS1 and RGS2 and fitted with the power-law model. We noticed that there are positive residuals around 0.52 KeV and 0.65 KeV as seen in Fig. 3. They are most likely associated with the emission lines of NVII and OVIII. The lines appear slightly broad and therefore they are represented as a gravitationally redshifted disc line (LAOR in XSPEC). Adding two LAOR lines to the power law model significantly improves the quality of the fit ($\Delta\chi^2/\Delta\nu = -111.4/-7$). The fit with PL+2LAOR model is shown in Fig. 2. The presence of the lines N and O is required at more than 10σ level of confidence as determined by their normalization.

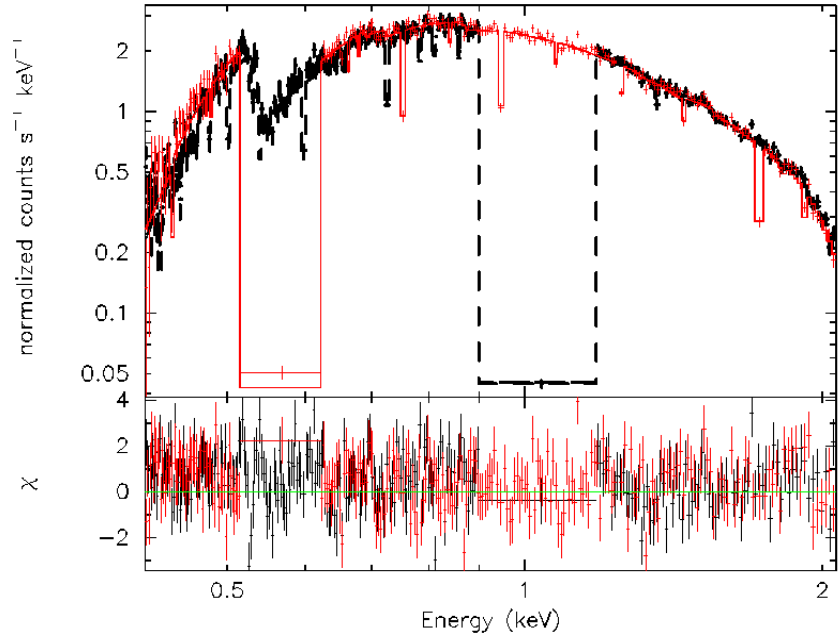

Fig. 2 Fit of a PL+2LAOR model to RGS1 and RGS2 combined spectrum. RGS1 is in black (dashed) and RGS2 (solid-line) is in red. The gaps in the RGS spectra are due to CCD failture.

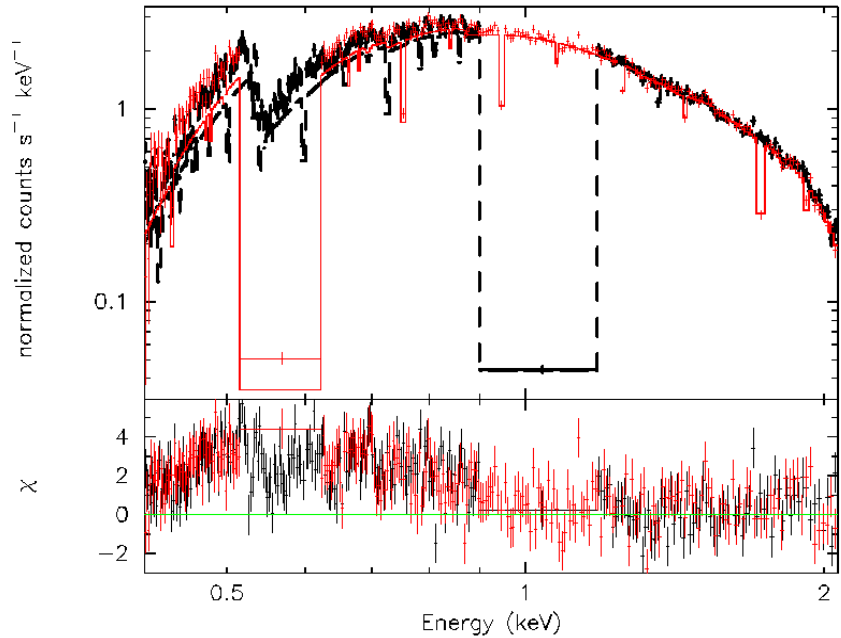

Fig. 3 Fit of a PL+2LAOR model to RGS1 and RGS2 combined spectrum with setting the normalization of the lines to zero shows positive residuals around 0.52 keV and 0.65 keV for the NVII and OVIII emission lines. RGS1(dashed) is in black and RGS2 (solid-line) is in red. The gaps in the RGS spectra are due to CCD failture.

Therefore, for all fits below the lines of NVII and OVII I are modeled by LAOR line profile. All the parameters of the three LAOR lines are linked except the line energies and the normalizations. We constrain the line energy for N and O to a range between 0.48 and 0.55 keV and 0.64 and 0.69 keV, respectively.

We initially fit the spectra with a power-law model (PL). For this model $\chi^2$ per degree of freedom (dof) was poor, typically 5606.42/4201 (see model 1 in Table 1). Adding the three LAOR lines for Fe, N and O to this model improves the quality of the fit, resulting in a reduced $\chi^2$ of 1.19 for 4192 dof (see model 2 in Table 1). The normalization of Fe, N, and O lines are more than $10\sigma$ different from zero and their equivalent width are $378.04_{-20.86}^{+61.91}$ eV, $13.8_{-5.20}^{+4.91}$ eV, and $15.7_{-4.94}^{+3.45}$ eV, respectively. To test whether a soft component is present in the spectra, we added a disc-blackbody component to the PL+3LAOR model. This gives a relatively large improvement in the fit ($\Delta\chi^2/\Delta\nu = -255.8/-2$; see model 3 in Table 1) with a column density $N_H$ of $0.33 \times 10^{22}$ cm$^{-2}$ and an inner disc temperature of $0.25 \pm 0.01$ KeV. The normalization of the lines Fe, N, and O are more than $10\sigma$ different from zero with an equivalent width of $185.26_{-44.23}^{+33.77}$ eV, $15.49_{-12.83}^{+11.11}$ eV, and $51.71_{-6.40}^{+24.48}$ eV, respectively. The presence of the DBB component is required at the $5\sigma$ level of confidence as determined by the DBB normalization so it is required by the data. Also, an F-test indicates that there is a probability of $10^{-48}$ that the improvement is only by chance.

Fig. 4 shows the best-fitting spectra. The absorbed flux (0.6–10.0 keV) is $3.60 \pm 0.03 \times 10^{-10}$ erg cm$^{-2}$ s$^{-1}$ where the flux of the DBB component is $1.71_{-0.35}^{+0.47} \times 10^{-11}$ erg cm$^{-2}$ s$^{-1}$ (0.6–10.0 keV) which contributes only 4.7% of the total flux.

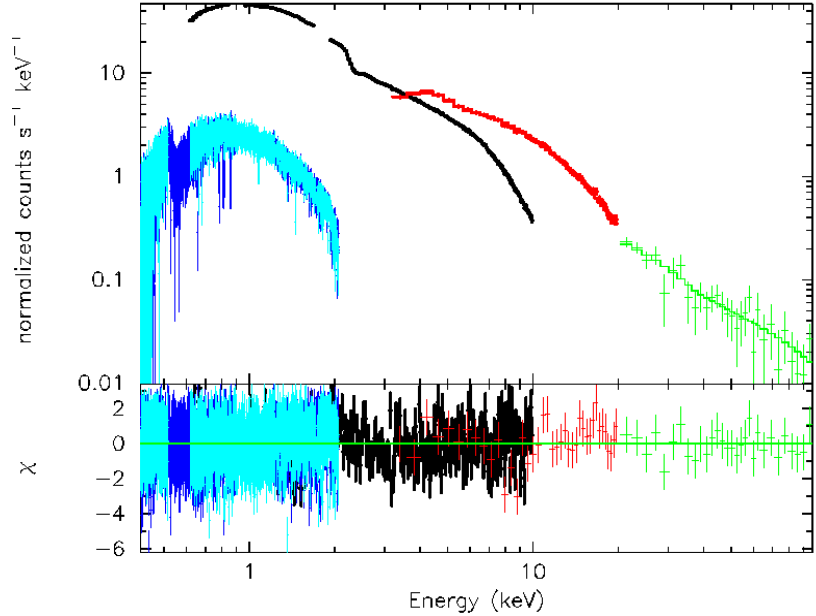

**Fig. 4** Broadband spectrum obtained from fitting PCA, HEXTE, RGS1,2, and PN spectra of Swift J1753.5-0127 with PL+DBB+3LAOR. RGS spectrum (0.4-2.1 keV) is shown in blue, RGS2 spectrum (0.4-2.1 keV) in blue-green, PN spectrum (0.6-10 keV) in black, PCA spectrum (3-20 keV) in red, and HEXTE spectrum (20-100 keV) in green.

It is believed that the hard component is due to inverse Compton scattering of seed photons from the accretion disc. Therefore, we can approximate the non-thermal emission with the Comptonization model (Titarchuk 1994) in our fits. The COMPTT model does not provide a good fit (reduced $\chi^2$ = 1.21 for 4199 dof, see model 4 in Table 2). When three LAOR lines are added to the model, the fits show a large improvement of $\Delta\chi^2/\Delta\nu = -276.9/-9$ (see model 5 in Table 2). The normalization of Fe, N, and O lines are more than 10$\sigma$ different from zero and their equivalent width are $119.61^{+37.23}_{-35.20}$ eV, $24.36^{+7.71}_{-8.13}$ eV, and $51.06^{+13.8}_{-6.74}$ eV, respectively. Next we fitted the spectra with COMPTT+3LAOR+DBB model (model 5) and tied the seed photon temperature ($kT_0$ in COMPTT) to the inner accretion disc temperature ($kT_{in}$ in DISKBB). Adding a disc black-body model improves the fit with a change in the goodness-of-fit of $\chi^2/\Delta\nu$ = -42/-1 (see Table 2). The fitting shows that the disc black-body normalization is more than 10$\sigma$ different from zero. In addition, an F-test indicates that there is a probability of $10^{-9}$ that the improvement is only by chance. Fig. 5 shows the best-fitting model. The Fe, N, and O lines are found to have an

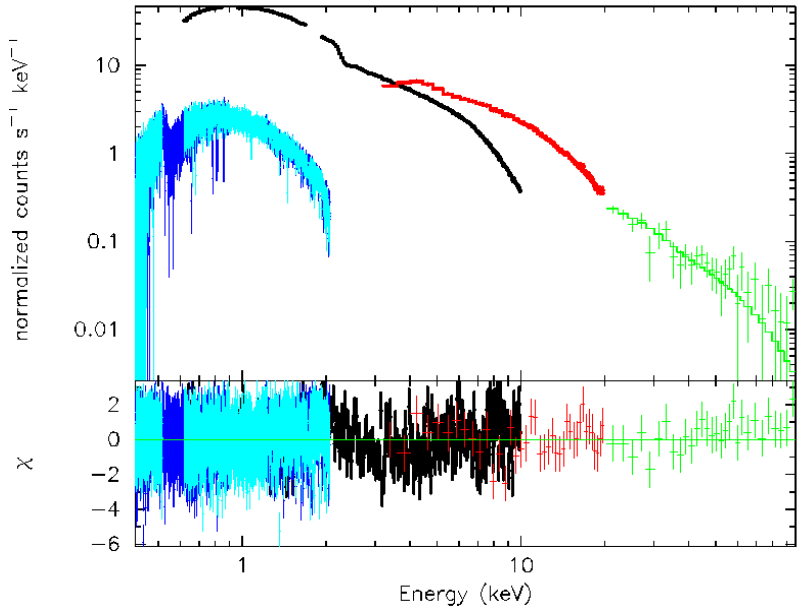

**Fig. 5** Broadband spectrum obtained from fitting PCA, HEXTE, RGS1,2, and PN spectra of Swift J1753.5-0127 with COMPTT+DBB+3LAOR. RGS spectrum (0.4-2.1 keV) is shown in blue, RGS2 spectrum (0.4-2.1 keV) in blue-green, PN spectrum (0.6-10 keV) in black, PCA spectrum (3-20 keV) in red, and HEXTE spectrum (20-100 keV) in green.

equivalent width of $93.02^{+21.39}_{-25.09}$ eV, $37.51^{+9.25}_{-3.62}$ eV, and $26.97^{+4.99}_{-3.70}$ eV, respectively and a normalization that is more than 10$\sigma$ different from zero. The absorbed flux is $3.60^{+0.04}_{-0.01} \times 10^{-10}$ erg cm$^{-2}$ s$^{-1}$ (0.6–10.0 KeV) whereas the disc contributes with $1.47 \pm 0.04 \times 10^{-11}$ erg cm$^{-2}$ s$^{-1}$ (0.6–10.0 KeV).

Finally, we fitted the spectra with a constant density ionized reflection model, REFLIONX (Ross & Fabian 2005) in XSPEC. This model describes the reflected spectrum from an optically-thick atmosphere of constant density illuminated by a power-law spectrum with an exponential cutoff energy fixed at 300 keV. We first fitted REFLIONX+PL model (model 7 in Table 3) to the spectra, where we fixed the iron abundances to solar and tied the power-law index of REFLIONX to that of the power-law component. This fit yields a reduced $\chi^2$ of 1.18 (see Table 3). To verify the significance of the disc component we fitted the spectra with REFLIONX+PL+DBB model (model 8 in Table 3) which results in a large change in the goodness-of-fit of $\Delta\chi^2/\Delta\nu = -105.8/-2$ with a disc temperature of $0.20 \pm 0.001$ keV and the flux of the DBB component is $7.89^{+1.17}_{-0.65} \times 10^{-12}$ erg cm$^{-2}$ s$^{-1}$ (0.6–10.0 keV). This contributes 2.2% from the absorbed flux ($3.60 \pm 0.04 \times 10^{-10}$ erg cm$^{-2}$ s$^{-1}$) in the range 0.6-10.0 keV. The disc component is significantly required at more than $10\sigma$ level of confidence as determined from their normalization. Also, an F-test indicates that there is a probability of $10^{-20}$ that the improvement is only by chance. The flux of the reflection component is $4.84^{+0.01}_{-0.40} \times 10^{-11}$ erg cm$^{-2}$ s$^{-1}$ (0.6–10.0 keV). We then convolved the reflection component with the Kdblur relativistic smearing kernel to blur the reflection continuum and the lines as they affected by relativistic effects from an accretion disc. This Kdblur convolution model (KDBLUR in XSPEC) is derived from the laor code and assumes a maximally spinning black hole (a = 0.998). For kdblur, we fixed the outer radius of disc parameter to $R_{max} = 400 R_g$ and let the others float freely. Modeling the continuum with PL+kdblur(REFLIONX) results in a reduced $\chi^2$ of 1.15 for 4196 dof. Table 4 summarizes the parameters that have been obtained fitting with this model. Adding a disc black-body component with a temperature 0.52 $\pm$ 0.05 keV to the model PL+kdblur(REFLIONX) provides a slight improvement of $\Delta\chi^2/\Delta\nu = -12.9/-2$. The thermal component is significantly required at $5\sigma$ level of confidence as estimated from the normalization of the DISKBB. An F-test indicates that there is a probability of $4 \times 10^{-3}$ that the improvement is only by chance. The best fit model is shown in Fig. 6. The disc component contributes about 1.7% of the 0.6-

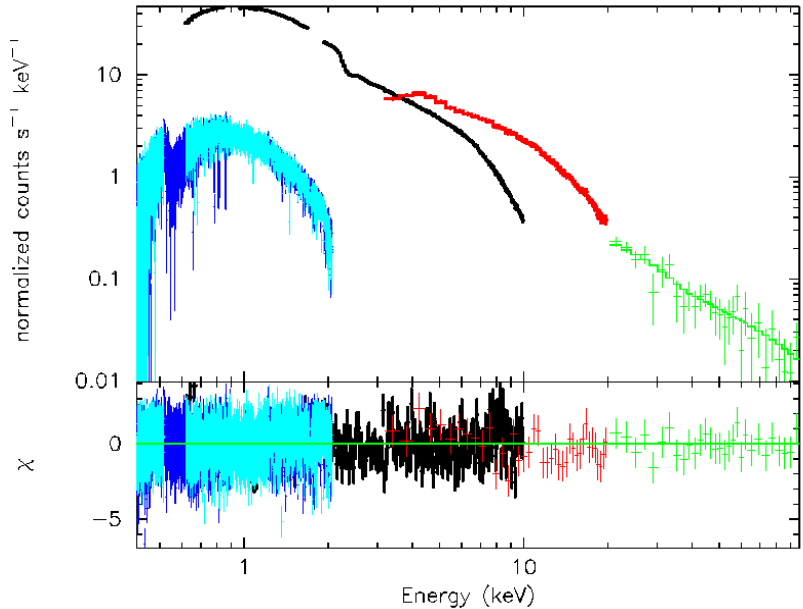

**Fig. 6** Broadband spectrum obtained from fitting PCA, HEXTE, RGS1,2, and PN spectra of Swift J1753.5-0127 with PL+DBB+Kdblur(REFLIONX) model. RGS spectrum (0.4-2.1 keV) is shown in blue, RGS2 spectrum (0.4-2.1 keV) in blue-green, PN spectrum (0.6-10 keV) in black, PCA spectrum (3-20 keV) in red, and HEXTE spectrum (20-100 keV) in green.

10.0 keV absorbed flux ($3.60^{+0.02}_{-0.05} \times 10^{-10}$ erg cm$^{-2}$ s$^{-1}$) whereas the reflection component contributes about 8.6%.

## 4. DISCUSSION AND CONCLUSIONS

Our main goal was to test the presence of the thermal component in the Swift J1753.5-0127 spectrum. Spectral fits of the 2006 *XMM-Newton*/*RXTE* spectrum with a range of continuum models that include and exclude thermal component reveal that the POWER-LAW, COMPTT, REFLIONX and REFLIONX convolved with Kdblur kernel models successfully fit the spectrum with the need of a disc black-body component. However, fitting the spectrum with REFLIONX convolved with Kdblur kernel model reveals that the disc black-body component is not strongly required compared with the other models (see the F-test probability). Thus, our results confirm the requirement of the thermal component by the data. However, the contribution of the disc is extremely low, from about 1.7% to 4.7% of the total flux in the range 0.6-10 keV, depending on the model we used. The fits of the soft emission arising from the accretion disc illustrated the presence of a cool inner disc with temperatures that range between 0.19-0.52 keV, depending on the model we used, during the low/hard state of the X-ray transient Swift J1753.5-0127. Previous analysis of the same XMM-Newton observations has reported the requirement of a soft thermal component (kT~0.2 keV) in the spectrum of Swift J1753.5-0127 (Miller et al. 2006a). Moreover the presence of such soft component was confirmed from Suzaku observations (Reynolds et al. 2009). In contrast, Hiemstra et al. 2009 could successfully fit the X-ray spectrum with a continuum model that does not a soft disc component. Additionally Tomsick et al. (2008) studied the broadband X-ray spectra of the X-ray transient GX 339-4 in the hard state using *Swift* and *RXTE* and found a thermal component with an inner disc temperature of 0.2 keV at 2.3% L$_{Edd}$ where, at 0.8% L$_{Edd}$ the spectrum is consistent with the presence of such a component, but not required with high confidence. The existence of a cool inner disc in black hole X-ray transient systems in the low/hard state was investigated by Liu et al. (2007) through the evaporation/condensation model. This model describes the exchange of energy and mass between the corona and the disc originating from thermal conduction including the effects of coronal cooling associated with the inverse Compton scattering of soft X-ray photons. They found that for a range of luminosities that depends on the value of the viscosity parameter weak condensation–fed inner disc can be present in the low/hard state of black hole transient systems. Such model was applied by Liu et al. (2007) to the two X-ray transients GX 339-4 and Swift J1753.5-0127 and they found that cool inner discs, contributing a small fraction of the total X-ray luminosity (<20%), can exist in the low/hard state. Therefore our result is in good accordance with this picture. Moreover, a soft emission in the low/hard state was also apparent in several black hole systems such as Cyg X-1 (Balucinska-Church et al. 1995; Di Salvo et al. 2001) and GRO J1655-40 (Brocksopp et al. 2006; Takahashi et al. 2008).

We confirm the presence of a broad iron emission line which was previously discovered by Hiemstra et al. (2009). We fitted the broad Fe Kα emission line in our spectrum using the LAOR relativistic line model, which provided a much better fit to our data than a Gaussian model. The line is significantly required by the data as indicated by the line normalization which is more than 10σ different from zero, in the best-fitting models. This is in consistent with a previous results obtained by Hiemstra et al. (2009) (3-14σ significant depending on the model used) and Reynolds et al. (2009) (>7σ significant). It is thought that the Fe line can be broadened as a consequence of a relativistic distortion, if it is emitted from an accretion disc that is close to the black hole (Fabian et al. 2000). As determined from the LAOR line profile in the best-fitting models the line equivalent width ranges between 93 and 185 eV, depending on the used model. Similarly, Hiemstra et al. (2009) determined the equivalent width to be 60-187 eV depending on the model used, while Reynolds et al. 2009 found an equivalent width of ∼70 ± 30 eV. Results from the best-fitting models suggest that the inner edge of the accretion disc extend down 1.7-7 Rg, which indicates non-truncated disc that remain at or close to the innermost circular orbit. The discrepancy which is found in the fitting with LAOR model is the inclination of the system. For the different best-fitting models, LAOR fits tend to give inclination values higher than 86° which conflict with the fact that there are no eclipses detected in the X-ray light curve of Swift J1753.5-0127. In the case of fitting the data with a smeared reflection model: Kdblur*REFLIONX, the inner radius determined is 10.9 Rg, which seems to suggest a disc truncated at small radius. This radius is larger than what we obtained from fits to the Fe Kα line using the LAOR profile (ranges between 1.7-7 $R_g$). In comparison, previous fits to the Fe Kα line, using a loar profile, yielded an inner disc radius varies from 5.5 $R_g$ to 14.6 $R_g$ for the best-fit models, although fitting with reflection models smeared by relativistic effects suggested much larger radius $R_{in}$ ∼ 250 $R_g$ which indicate that the inner disc radius is strongly depends on the continuum model used (Hiemstra et al. 2009). Estimates of the inner disc radius inferred from the laor line fits done by Reynolds et al. (2009) showed a disc truncated at small radii ($R_{in}$ ∼ 10-20 $R_g$), on the other hand fitting with a simple diskbb+powerlaw model and a reflection model revealed a disc consistent with ISCO ($R_{in} \leq 6R_g$) for certain values of both the column density and inclination. We also found that the contribution of the reflection component is about 8.6%, which implies that the other 89.7% should appear as reprocessed radiation in the disc.

The most physically realistic model used, the smeared reflection model, reveals a truncated disc at a small radius ($R_{in}$ ∼ 11 $R_g$). Then our results suggest the possibility for a truncated accretion disc at small radii in the low/hard state. This is in agreement with the results of previous analysis by Hiemstra et al. (2009). They also suggested a disc truncated at much larger radius $R_{in}$ ∼250 Rg in the case of fitting with REFLION model convolved with KERRCONV. However, Miller et al. (2006a) revealed a cool accretion disc that extends close to the ISCO. A spectral fit, including a power law and a self-consistent component, was employed by Reis et al. 2009; they found an inner disc radius of about 3.1Rg. Similar to this finding, an observation of GX 339-4 suggested a disc that remains at or near the innermost stable circular orbit in the low/hard state (Miller et al. 2006b). The most common model suggested for describing the truncated disc picture was proposed by Esin, McClintock & Narayan (1997). In this model, the low/hard state is identified by a radially truncated geometrically thin, an optically thick accretion disc at some radius, transition radius $r_{tr}$, and an inner advection dominated accretion flow (ADAF), $r < r_{tr}$, which radiates less efficiently. This transition radius is a function of the mass accretion rate .The hard X-rays are produced by the inverse Compton of soft photons. This model is appropriate in explaining our spectrum. Moreover, Wilkinson & Uttley 2009 presented a new spectral analysis technique, the 'covariance spectrum', to study the variability spectra from XMM-Newton observations of Swift J1753.5-0127 and GX 339-4. They found that disc blackbody emission contributes significantly to the X-ray variability spectra. They interpret the extra blackbody variability seen on longer time-scales to be intrinsic to the accretion disc. They suggested that such variations are likely to be responsible for the low frequency Lorentzian component in the hard state power spectral density function in both sources. This implies a disc truncation radius < 20 $R_g$. This is in convenient with our estimate of the disc inner radius inferred from the smeared reflection model which indicates a truncated disc at small radius ($R_{in}$ ~ 11$R_g$). A comparative spectral-timing study of XMM-Newton data from the source in a bright 2009 hard state with that from the fainter 2006 hard state was done recently by Cassatella et al. (2012). They found that in the bright 2009 hard state the disc variability below 0.6 keV becomes incoherent with the power-law and higher-energy disc emission at frequencies below 0.5 Hz which is in contrast with the coherent variations seen in the 2006 data. Furthermore, they found that the characteristic frequencies are higher at higher luminosity.

We report the discovery of broad emission lines of NVII and OVIII at ~ 0.52 KeV and 0.65 KeV, respectively in the RGS spectrum of Swift J1753.5-0127. This feature is interpreted as caused by reflection of X-ray photons off accretion disc enriched with neon and oxygen elements. It is considered that OVIII line comes second in strength after the Fe Kα line in the case of an accretion disc with solar abundances (Madej & Jonker 2011). We fitted these lines with laor profile since they are thought to be affected by the strong gravity close to the black hole. We found that the NVII line is significantly required in the data as determined by its normalization for all models we used. However, the OVIII line is significantly required in the Power-law and Comptt models, but is not significantly required for the rest of the models. In fact, more analysis is necessary to confirm the presence of NVII and OVIII lines. A similar OVIII line was detected in the ultracompact X-ray binary 4U 0614+09 (Madej et al. 2010) and 4U 1543-624 (Madej & Jonker 2011).

Another feature observed in modeling the X-ray continuum with different components is column density $N_H$ showing a large variability during the fits, has a range of 2.28-3.30 (in the units of $10^{21}$ cm$^{-2}$) depending on the model used. The best-fitting spectra show that the continuum is well modeled with a moderate galactic absorption $N_H$ = 2.75-3.30 (in the units of $10^{21}$ cm$^{-2}$). It has been previously calculated to be $N_H$ = 2.3 ×$10^{21}$ cm$^{-2}$ from XMM-Newton observations (Miller et al. 2006a), while Cadolle Bel et al. (2007) obtained $N_H$ = 1.97 ± 0.23×$10^{21}$ cm$^{-2}$ from optical measurements. In addition, Reynolds et al. 2009 determined the best fit column density as measured by Suzaku to be $N_H$ = 1.8 ± 0.1×$10^{21}$ cm$^{-2}$. However, they noted that the measured column density depends on the model used as for the comptonization corona model the best fit value was $N_H$ = 3.1 ± 0.1×$10^{21}$ cm$^{-2}$.

Table 1: Fit results using POWER-LAW model

| Parameter | PL (model 1) | PL + 3LAOR (model 2) | PL + DBB + 3LAOR (model 3) |
|---|---|---|---|
| $N_H$ ($10^{21}$ cm$^{-2}$) (phabs) | $2.39 \pm 0.01$ | $2.67 \pm 0.01$ | $3.30 \pm 0.07$ |
| $\Gamma_{PL}$ (PL) | $1.63 \pm 0.00$ | $1.68 \pm 0.00$ | $1.64 \pm 0.01$ |
| $N_{PL}$ ($10^{-2}$) (PL) | $6.10 \pm 0.01$ | $6.06 \pm 0.00$ | $5.73 \pm 0.05$ |
| $E_L$ (keV) (Fe LAOR) | - | $6.40^{+0.07}_{-6.40}$ | $7.00^{+0.00}_{-0.08}$ |
| $E_L$ (keV) (N LAOR) | - | $0.48 \pm 0.00$ | $0.48 \pm 0.00$ |
| $E_L$ (keV) (O LAOR) | - | $0.69 \pm 0.00$ | $0.67 \pm 0.01$ |
| Index ($\alpha$) (LAOR) | - | $3.94 \pm 0.07$ | $3.24^{+0.11}_{-0.06}$ |
| $R_{in}$ ($R_g$) (LAOR) | - | $1.31 \pm 0.05$ | $1.66^{+0.09}_{-0.06}$ |
| i (deg) (LAOR) | - | $90.0^{+0.00}_{-0.18}$ | $90.0^{+0.00}_{-0.30}$ |
| $N_L$ ($10^{-4}$) (Fe LAOR) | - | $9.35 \pm 0.27$ | $4.42^{+0.46}_{-0.29}$ |
| $N_L$ ($10^{-3}$) (N LAOR) | - | $2.47^{+0.16}_{-0.19}$ | $14.44^{+1.87}_{-1.36}$ |
| $N_L$ ($10^{-3}$) (O LAOR) | - | $1.59^{+0.12}_{-0.07}$ | $2.36^{+0.40}_{-0.18}$ |
| $KT_{in}$ (keV) (DBB) | - | - | $0.25 \pm 0.01$ |
| $N_D$ (DBB) | - | - | $535.7^{+146.1}_{-108.2}$ |
| $\chi^2/\nu$ | 5606.4/4201(1.33) | 4989.5/4192(1.19) | 4733.7/4190(1.13) |

Notes: The phabs parameter is the equivalent hydrogen column, $N_H$. The power-law parameters are the photon index, $\Gamma$, and the normalization, $N_{PL}$ (in units of photons KeV$^{-1}$cm$^{-2}$s$^{-1}$ at 1 KeV). The parameters of LAOR are the line energy, $E_L$, emissivity index, index, inner radius, $R_{in}$ (in units of $GM/c^2$), inclination, i, and normalization, $N_L$ (in units of photons cm$^{-2}$s$^{-1}$). The diskbb parameters are the inner disc temperature, $KT_{in}$, and the normalization, $N_D$.

Table 2 Fit results using COMPTT model.

| Parameter | COMPTT (model 4) | COMPTT +3LAOR (model 5) | COMPTT+3LAOR+DBB (model 6) |
|---|---|---|---|
| $N_H$ ($10^{21}$ cm$^{-2}$) (phabs) | 2.28 ± 0.02 | 2.23 ± 0.01 | 2.83 ± 0.02 |
| $KT_o$ (keV) (compTT) | 0.13 ± 0.00 | 0.17 ± 0.00 | 0.19 ± 0.00 |
| kT (keV) (compTT) | $7.67^{+0.47}_{-0.42}$ | $12.12^{+0.02}_{-0.06}$ | $13.77^{+0.03}_{-0.12}$ |
| τ (compTT) | 4.19 ± 0.02 | $3.22^{+0.02}_{-0.19}$ | 2.99 ± 0.00 |
| $N_C$ ($10^{-2}$) (compTT) | 2.73 ± 0.18 | $1.47^{+0.18}_{-0.01}$ | 1.20 ± 0.00 |
| $E_L$ (keV) (Fe LAOR) | - | $6.75^{+0.00}_{-0.04}$ | $7.00^{+0.00}_{-0.04}$ |
| $E_L$ (keV) (N LAOR) | - | $0.49^{+0.02}_{-0.00}$ | 0.48 ± 0.00 |
| $E_L$ (keV) (O LAOR) | - | 0.68 ± 0.00 | 0.65 ± 0.00 |
| Index (α) (LAOR) | - | 2.49 ± 0.06 | 2.13 ± 0.05 |
| $R_{in}$ ($R_g$) (LAOR) | - | $1.95^{+0.12}_{-0.08}$ | $6.99^{+0.28}_{-0.64}$ |
| i (deg) (LAOR) | - | $88.4^{+0.75}_{-0.51}$ | $86.11^{+0.19}_{-0.46}$ |
| $N_L$ ($10^{-4}$) (Fe LAOR) | - | $3.01^{+0.22}_{-0.18}$ | 2.15 ± 0.19 |
| $N_L$ ($10^{-3}$) (N LAOR) | - | $7.59^{+0.59}_{-0.29}$ | $8.94^{+0.44}_{-0.38}$ |
| $N_L$ ($10^{-3}$) (O LAOR) | - | 2.53 ± 0.09 | $3.69^{+0.10}_{-0.15}$ |
| $KT_{in}$ (keV) (DBB) | - | - | 0.19 ± 0.00 |
| $N_D$ (DBB) | - | - | $2196.25^{+20.75}_{-19.03}$ |
| $\chi^2/\nu$ | 5083.2/4199(1.21) | 4806.3/4190(1.15) | 4764.3/4189(1.14) |

Notes: The comptt parameters are the input seed photon temperature, $KT_o$, plasma temperature, KT, plasma optical depth, τ, and normalization, $N_C$.

Table 3 Fit results using REFLIONX model

| Parameter | PL+REFLIONX (model 7) | PL+REFLIONX+DBB (model 8) |
|---|---|---|
| $N_H$ ($10^{21}$ cm$^{-2}$) (phabs) | $2.64 \pm 0.01$ | $2.93^{+0.07}_{-0.04}$ |
| $\Gamma$ (PL) | $1.57 \pm 0.01$ | $1.59 \pm 0.01$ |
| $N_{PL}$ ($10^{-2}$) (PL) | $4.88 \pm 0.08$ | $5.15^{+0.06}_{-0.12}$ |
| $\Gamma_R$ (Reflionx) | $1.57 \pm 0.01$ | $1.59 \pm 0.01$ |
| $\zeta_i$ (erg cm/s) (Reflionx) | $1579.1^{+87.40}_{-10420}$ | $1667.2^{+20420}_{-78.12}$ |
| $N_R$ ($10^{-7}$) (Reflionx) | $1.91 \pm 0.09$ | $1.37^{+0.16}_{-0.11}$ |
| $KT_{in}$ (keV) (DBB) | - | $0.20 \pm 0.001$ |
| $N_D$ (DBB) | - | $942.7^{+14040}_{-77.01}$ |
| $\chi^2/\nu$ | 4962.9/4199(1.18) | 4857.1/4197(1.16) |

Notes: The Reflionx parameters are the photon index, $\Gamma_R$, ionization parameter, $\zeta_i$, and normalization, $N_R$.

Table 4 Fit results using REFLIONX model convolved with Kdblur smearing kernel

| Parameter | PL+kdblur(REFLIONX) (model 9) | PL+kdblur(REFLIONX)+DBB (model 10) |
|---|---|---|
| $N_H$ ($10^{21}$ cm$^{-2}$) (phabs) | $2.76 \pm 0.02$ | $2.75 \pm 0.02$ |
| $\Gamma$ (PL) | $1.60 \pm 0.00$ | $1.59 \pm 0.01$ |
| $N_{PL}$ ($10^{-2}$) (PL) | $5.34 \pm 0.05$ | $5.38 \pm 0.08$ |
| Index (Kdblur) | $2.46^{+0.09}_{-0.16}$ | $10^{+0.00}_{-5.43}$ |
| $R_{in}$ ($R_g$) (Kdblur) | $1.31^{+0.31}_{-1.31}$ | $10.94^{+2.31}_{-4.11}$ |
| i (deg) (Kdblur) | $90.0^{+0.00}_{-2.13}$ | $86.97^{+0.00}_{-0.51}$ |
| $\Gamma_R$ (Reflionx) | $1.60 \pm 0.00$ | $1.59 \pm 0.01$ |
| $\zeta_i$ (erg cm/s) (Reflionx) | $722.5^{+51.32}_{-41.43}$ | $573.4^{+24.90}_{-27.80}$ |
| $N_R$ ($10^{-7}$) (Reflionx) | $4.08^{+0.36}_{-0.24}$ | $4.91^{+0.43}_{-0.36}$ |
| $KT_{in}$ (keV) (DBB) | - | $0.52 \pm 0.05$ |
| $N_D$ (DBB) | - | $5.26^{+2.36}_{-1.10}$ |
| $\chi^2/\nu$ | 4835.5/4196(1.15) | 4822.6/4194(1.15) |

Notes: The Kdblur parameters are the index, inner radius, $R_{in}$ (in units of GM/c$^2$), and inclination, i.